\documentclass[12pt, conference]{IEEEtran}

\usepackage{cite}

\ifCLASSINFOpdf
  \usepackage[pdftex]{graphicx}
  \DeclareGraphicsExtensions{.pdf,.jpeg,.png}
\else
\fi

\usepackage{amsmath}
\usepackage{amssymb}

\usepackage{array}

\ifCLASSOPTIONcompsoc
 \usepackage[caption=false,font=normalsize,labelfont=sf,textfont=sf]{subfig}
\else
 \usepackage[caption=false,font=footnotesize]{subfig}
\fi

\usepackage{stfloats}

\usepackage{url}

\begin{document}
\title{PHz Electronic Device Design and Simulation for Waveguide-Integrated Carrier-Envelope Phase Detection}






\author{\IEEEauthorblockN{
Dario Cattozzo Mor,\IEEEauthorrefmark{1}$^,$\IEEEauthorrefmark{5}
Yujia Yang,\IEEEauthorrefmark{1}
Felix Ritzkowsky,\IEEEauthorrefmark{1}$^,$\IEEEauthorrefmark{2}
Franz X. K\"artner,\IEEEauthorrefmark{2}\\
Karl K. Berggren,\IEEEauthorrefmark{1} 
Neetesh Kumar Singh\IEEEauthorrefmark{2}, and
Phillip D. Keathley\IEEEauthorrefmark{1}
}\\
\IEEEauthorblockA{\IEEEauthorrefmark{1}Research Laboratory of Electronics, Massachusetts Institute of Technology,\\77 Massachusetts Avenue, Cambridge, MA 02139, USA.\\
e-mail: pdkeat2@mit.edu}
\vspace{0.1cm}
\IEEEauthorblockA{\IEEEauthorrefmark{5}Department of Electronics and Telecommunications, Polytechnic University of Turin,\\Corso Duca degli Abruzzi 24, Turin 10129, Italy}
\vspace{0.1cm}
\IEEEauthorblockA{\IEEEauthorrefmark{2}Deutsches Elektronen Synchrotron (DESY) $\&$ Center for Free-Electron Laser Science $\&$ \\ University of Hamburg, Notkestraße 85, Hamburg 22607, Germany.\\ e-mail: neetesh.singh@cfel.de}
}

\maketitle

\begin{center}
    \large{September, 16$^{th}$ 2021}\\
\end{center}

\vspace{0.1cm}
\begin{abstract}

Carrier-envelope phase (CEP) detection of ultrashort optical pulses and low-energy waveform field sampling have recently been demonstrated using direct time-domain methods that exploit optical-field photoemission from plasmonic nanoantennas. These devices make for compact and integratable solid-state detectors operating at optical frequency that work in ambient conditions and require minute pulse energies (picojoule-level).  Applications include frequency-comb stabilization, visible to near-infrared time-domain spectroscopy, compact tools for attosecond science and metrology and, due to the high electronic switching speeds, petahertz-scale information processing.
However, these devices have been driven by free-space optical waveforms and their implementation within integrated photonic platforms has yet to be demonstrated.
In this work, we design and simulate fully-integrated plasmonic bow-tie nanoantennas coupled to a Si$_3$N$_4$-core waveguide for CEP detection. We find that when coupled to realistic on-chip, few-cycle supercontinuum sources, these devices are suitable for direct time-domain CEP detection within integrated photonic platforms.  We estimate a signal-to-noise ratio of 30 dB at 50 kHz resolution bandwidth.
We address technical details, such as the tuning of the nanoantennas plasmonic resonance and the waveform's CEP slippage in the waveguide. Moreover, we evaluate power losses due to absorption and scattering and we study the device sensitivity to pulse duration and pulse peak field intensity.
Our results provide the basis for future design and fabrication of time-domain CEP detectors and allow for the development of fully-integrated attosecond science applications, frequency-comb stabilization and light-wave-based PHz electronics.

\end{abstract}

\ifCLASSOPTIONpeerreview
\begin{center} \bfseries EDICS Category: 3-BBND \end{center}
\fi
%
\IEEEpeerreviewmaketitle

\section{Introduction}

Ultrafast, strong-field light-matter interactions between high-intensity, few-cycle pulses and nanoscale antenna structures enable PHz processing of ultrafast optical waveforms \cite{paper:tips_3, QNN:yujia, paper:mina, paper:dombi, paper:schultze_dielectrics, paper:paasch, paper:racz, paper:tips_6, paper:rybka, QNN:putnam, paper:karnetzky, QNN:vanishing, paper:zimmermann, paper:sederberg, paper:ludwig}.
In these interactions, strong electric fields (tens of GV/m) generate photocurrents by optical-field-emission processes. With devices exploiting these currents,
time-domain carrier-envelope phase (CEP) detection for ultrashort pulses \cite{QNN:yujia, paper:paasch, paper:rybka, QNN:putnam} and field sampling of low-energy waveforms \cite{paper:mina} have been demonstrated.

CEP is the feature of an optical pulse that describes the offset between the peak of the electric field carrier wave and the peak of the pulse envelope. For example, consider an electric field waveform defined as $F(t) = F_0(t)\cos(\omega t + \varphi _{\text{CE}})$.  The carrier angular frequency is defined by $\omega$, the pulse envelope by $F_0(t)$, and the carrier-envelope phase by $\varphi_{\text{CE}}$. CEP defines the exact shape of the optical electric field waveform, and its variation in time impacts the frequency offset of broadband frequency-comb sources.  As such, CEP is of critical importance to few-cycle light–matter interactions~\cite{paper:kling, paper:bergues, paper:debohan, paper:christov} and frequency metrology~\cite{paper:metrology, paper:freq_combs,paper:freq_combs_singh}, and for emerging applications in ultrafast information processing using lightwave-based PHz electronics \cite{paper:perspect_petahz_electr, paper:krausz_stockman_atto_metr, paper:goulielmakis, paper:lee_petahertz_1, paper:lee_petahertz_2}.

Given the rapid progress in the development of integrated few-cycle laser sources within waveguides \cite{paper:SiN_Carlson, ceo_stab_Klenner:16, ceo_stab_Okawachi:18, paper:integr_sources, paper:integr_sources_2}, it is important that CEP detection devices also be integrated into monolithic photonic platforms.
Currently, CEP detection is mainly achieved with frequency-domain self-referencing techniques based on $f$-$2f$ interferometers \cite{paper:yujia_2, paper:yujia_3, paper:yujia_4}.  Recently, the integration of these frequency-domain techniques within photonic waveguides has been demonstrated \cite{paper:okawachi_gaeta_LN_integr_combs, paper:brasch_kippenberg_photnic_combs, paper:carlson_diddams_SiN_combs,paper:ceo_stab_diddams, paper:freq_combs_singh}. These techniques enable compact mm-scale, waveguide-integrated CEP-detection, but use $\chi^{(2)}$ materials that are difficult to incorporate into traditional integrated photonic fabrication processes.

\begin{figure*}[!h]
\centering
\includegraphics[width=1\linewidth]{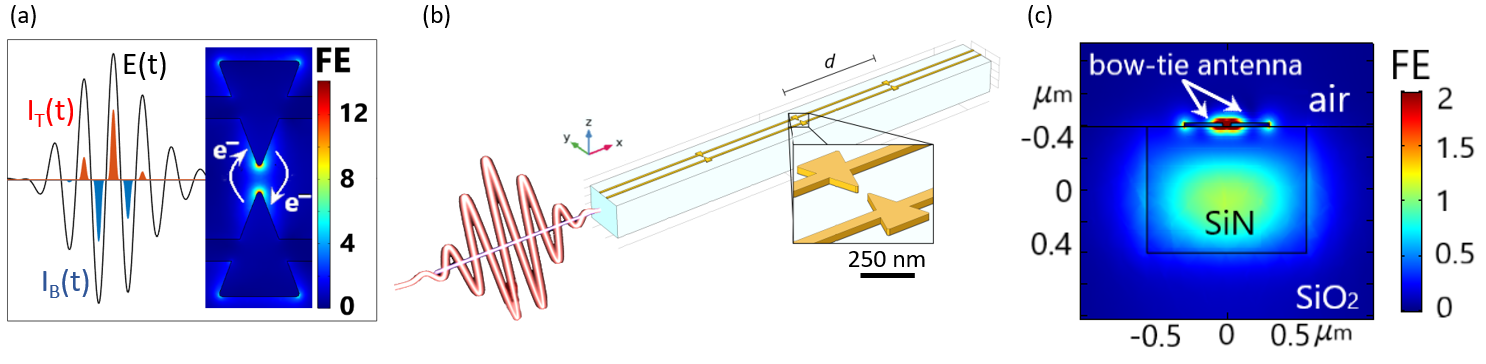}
\caption{(a) Field waveform of the enhanced pulse at the device first antenna tips and resulting photocurrent bursts. The current from the top antenna is shown in red, and the current from the bottom antenna is shown in blue. The inset shows the field enhancement at one of the tips in the first antenna structure.
(b) To-scale schematic of electrically-connected bow-tie CEP detectors integrated onto a Si$_3$N$_4$ waveguide. The spacing $d$ was chosen such that the carrier-envelope phase is the same at each antenna. The inset shows the bow-tie nanoantenna configuration. (c) Cross-sectional field distribution shown at the center of a nanoantenna device for $\lambda = 1.55\,\mu$m. X-axis is the waveguide width and y-axis the height. The field distribution is normalized relative to the injected pulse at the start of the waveguide. Note that the color scale is saturated to clearly visualize the waveguide mode. The field at the antennas tips is approximately $10\times$ the field in the center of the waveguide core. 
}
\label{fig_device}
\end{figure*}

An alternative approach has been recently demonstrated that uses optical-field emission from plasmonic nanoantennas which exhibits a CEP-sensitive optoelectronic response~\cite{QNN:yujia, paper:rybka, paper:ludwig}.  These detectors operate directly in the time-domain, and are similar to large-scale detectors that rely on strong-field photoemission from gases for shot-to-shot CEP tagging \cite{paper:wittman_ati, paper:paulus_ati}.  Due to the nanoantenna's ability to efficiently collect light, they can be scaled to very compact foot-prints, of the order of hundreds of nm to several $\mu$m, and generate large near-field enhancements drastically reducing the required pulse energies needed for operation relative to similar gas-phase techniques (down to tens to hundreds of pJ). Furthermore, the devices work in ambient conditions, without the need for vacuum packaging, and the CEP detection happens completely within the antenna structures, removing the need for a separate photodetection stage.  
While previous demonstrations of these PHz electronic nanoantenna detectors have been driven by free-space optical waveforms, their integration within integrated photonics has yet to be explored.

In this work, we design and simulate electrically-connected, waveguide-integrated linear nanoantenna arrays that are coupled to a Si$_3$N$_4$ waveguide for operation within monolithic integrated photonic platforms. 

The integration of plasmonic nanoantennas and photonic waveguides has been demonstrated before for the study of scattering properties of gold nanoantennas arrays \cite{paper:plasmonic_wg, paper:plasmonic_wg_3}, and here we specifically study the ability for these waveguide-integrated structures to detect the CEP of few-cycle supercontinuum optical waveforms that are injected into the waveguide.
In section \ref{work_princ} we overview the design and present the detector's working principle. We then address the technical challenges of the integration on waveguide, such as the tuning of the nanoantennas' plasmonic resonance and the waveform's CEP slippage in the waveguide.
In section \ref{results}, we calculate the emitted photocurrent, CEP-sensitivity and CEP-sensitive current SNR for a total of nine antennas coupled to the waveguide when it is injected with realistic supercontinuum pulses When driven by these supercontinuum pulses, we calculate an expected SNR of 30 dB at a resolution bandwidth (RBW) of 50 kHz, which is suitable for CEP detection and comparable to the performance of state-of-the-art self-referencing techniques that have been recently published ($\approx 30$ dB at RBW of tens to hundreds of kHz; see Refs.~\cite{paper:okawachi_gaeta_LN_integr_combs, ceo_stab_Klenner:16, ceo_stab_Okawachi:18}).

\section{Model setup and device design}
\label{work_princ}

The inset of Figure \ref{fig_device}(a) shows a single bow-tie nanoantenna CEP detector. Each nanoantenna consists of a pair of gold nanotriangles. For the simulations here we used a fixed gap width of 50 nm as this is similar to the gaps used in past work investigating similar devices in free-space~\cite{QNN:yujia}.  These bow-tie devices are electrically interconnected by gold nanowires shown in the unit cell in Figure~\ref{fig_device}(a), allowing them to be connected together along the surface of a waveguide as shown in Figure~\ref{fig_device}(b). Each nanoantenna behaves as a photoelectron-tunnelling device when driven by few-cycle waveforms with adequate peak intensity~\cite{QNN:yujia, paper:rybka}. The two nanotriangles are the cathode and the anode for electron emission and collection, respectively.

Any generated net photocurrent is sensitive to changes in the CEP. Consider that an ultrafast optical waveform illuminates the bow-tie nanoantenna.  Due to plasmonic resonance and the localization of the fields near the few-nanometer tips, the electric fields of the waveform are significantly enhanced. If enhanced to peak intensities such that the Keldysh parameter $\gamma \lesssim 1$ for the given nanoantenna material \cite{paper:keldysh, paper:yalunin}, optical-field-driven tunneling emission results between the two antennas tips of each bow-tie nanoantenna in the form of sub-optical-cycle current bursts \cite{QNN:putnam, QNN:vanishing, QNN:yujia}.  For illustration, we show the enhanced fields along with the simulated peak currents at the tip apices in Figure~\ref{fig_device}(a).  Photoemission occurs from either triangle tip depending on field polarity and the current emitted by the top nanotriangle is shown with a positive sign in red, while the current from the bottom is shown as negative in blue. 
In the inset of Figure \ref{fig_device}(a), the FE at one bow-tie nanoantenna is plotted.

For different CEP values of the incident waveform, different photoemission patterns can be observed and so the net emitted current, given by the integration in time of signals from both nanoantennas tips, varies. In particular, it shows a sinusoidal dependence on CEP and the amplitude of the oscillation is defined as CEP-sensitive current.
This can be expressed mathematically with the notation, similar to that used in ref. \cite{QNN:yujia}, which is: $I = I_T - I_B$, where $I_T$ is the photocurrent emitted by the top triangles tips towards the bottom ones, $I_B$ the one emitted by the bottom ones towards the top ones and $I$ is the net photocurrent. $I_T$ and $I_B$ can be expressed as:
\begin{equation}
    I_T \approx I_{0,T} + |I_{1,T}|\cos(\varphi_{\text{CE}} + \angle  I_{1,T}) 
    \label{eq:IL_IR}
\end{equation}
\vspace{-0.5cm}
\begin{equation}
    I_B \approx I_{0,B} - |I_{1,B}|\cos(\varphi_{\text{CE}} + \angle  I_{1,B})
    \label{eq:IL_IR_2}
\end{equation}
where $I_0$ is the total average photocurrent and $I_1$ the complex amplitude of the first harmonic of the CEP-sensitive photocurrent. For perfectly symmetric nanoantennas, as in our case, $|I_{1,T}| = |I_{1,B}|$, so $I = |I_{cep}|\cos(\varphi_{\text{CE}} + \angle  I_{1,T})$, with $|I_{cep}|$ the complex amplitude of the first harmonic of the total CEP-sensitive photocurrent and $|I_{cep}| = 2|I_{1,T}|$.


In this work, individual bow-ties are connected together for signal integration and directly placed onto the surface of a Si$_3$N$_4$ waveguide, as shown in Figure 1(b).
The waveguide was simulated as having a Si$_3$N$_4$ core with rectangular cross-section of $1\,\mu \text{m} \times 0.8\,\mu \text{m}$, a SiO$_2$ bottom cladding up to the top surface of the core and the antennas are exposed to air.

For developing the precise detector geometry for CEP detection, we simulated the electromagnetic response using a finite element method electromagnetic simulation (\textit{COMSOL Multiphysics}) over injected wavelengths from 1$\,\mu$m to 2.5$\,\mu$m.
In the initial design phase, we modeled devices consisting of three nanoantennas to decrease the simulation time for each tested geometry.  Then, once the basic nanoantenna geometry and spacings were chosen, we extend the number of antennas up to 9 to examine losses and impact on the CEP-sensitive photocurrent generated as a function of device length.  

\begin{figure}[!h]
  \centering
  \includegraphics[width=0.9\linewidth]{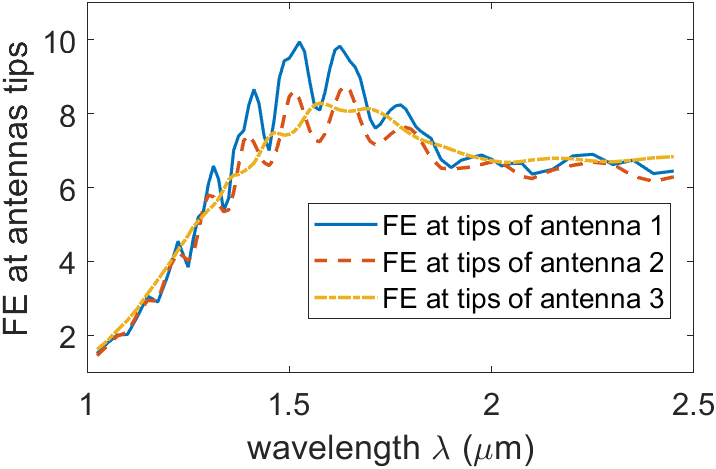}
\caption{
Calculated spectral field enhancement (FE) response at the tip near the nano-gap of each nanoantenna. The overall peak in FE results from the excitation of localized plasmonic resonances and it is centered at $\lambda = 1.55\,\mu$m, as by design. The weak modulation results from the excitation of cavity resonances in the regions between consecutive antennas. Assumed spacing between antennas $d=4.03~\mu$m. 
}
\label{fig_FE}
\end{figure}

Figure \ref{fig_device}(c) shows the cross-sectional field distribution at the center of a bow-tie device, for $\lambda = 1.55\,\mu$m. Due to plasmonic excitation, the field at the tips is enhanced up to approximately $10\times$ the field in the waveguide core (note that the FE color scale is saturated in the antenna to permit the reader to visualize the waveguide mode). This field enhancement value is lower than what was obtained in other works on free-space illuminated nanoantennas, such as in  \cite{QNN:yujia}, due to the fact that here the antennas are evanescently coupled to the waveguide and a smaller portion of the total input power is delivered to them.  However, this reduced field enhancement is largely compensated by the fact that: (1) the incident mode is highly confined by the waveguide structure; and (2) the passing radiation can further be detected by devices placed downstream.  

We designed the plasmonic nanoantenna to be resonant at $\lambda = 1.55\,\mu$m, the central wavelength of the driving pulses, for optimal enhancement.
We found that this target can be achieved by using a nanoantenna having triangle height of $\approx$ 250~nm and a base of $\approx$ 190~nm placed on top of the waveguide. The apices were rounded with 5 nm radius of curvature.
The connecting gold nanowires were taken to be 40 nm wide and the whole gold structure (connecting wires and nanotriangles) was 20 nm thick. A 2-nm-thick Chromium adhesion layer was included between the waveguide core and the gold nanostructures.

Figure \ref{fig_FE} shows the resulting spectral field enhancement (FE) response at the tip near the nano-gap of each antenna. The FE was defined as the field averaged over the tip surface, divided by a field reference value $E_0$, which is the maximum field in the center of the waveguide core, at the entrance of the waveguide. We call this value the injected pulse peak field.
The overall peak in FE results from the excitation of localized plasmonic resonances and it is centered at $\lambda = 1.55\,\mu$m, as by design.
The weak modulation of FE results from the excitation of cavity resonances in the regions between consecutive antennas. 

Next, we address the problem of CEP changes that are induced by propagation along the waveguide and interaction with the nanoantenna structures and connecting wires.  These CEP changes during propagation are due to a difference in the effective phase and group velocities that results from index dispersion of the propagating waveguide modes. Since this phase and group velocity difference creates a continuous shift in the CEP of the passing pulse as a function of propagation distance, we refer to it as CEP ``slippage''.  This CEP slippage is a critical concern for waveguide-integrated CEP detection since the signal emitted from each antenna is added to form a total signal, meaning that different CEP values at each antenna would result in a reduction of the net CEP-sensitive current accumulated from the entire device per incident pulse.  
We accounted for the CEP slippage in the waveguide by positioning the antennas at a certain distance from one another such that the CEP slips by an integer multiple of $2\pi$ between each antenna.  In this configuration, the signal emitted from each antenna adds in phase and contains information on the CEP of the injected waveform.

\begin{figure*}[!t]
  \centering
  \includegraphics[width=1\linewidth]{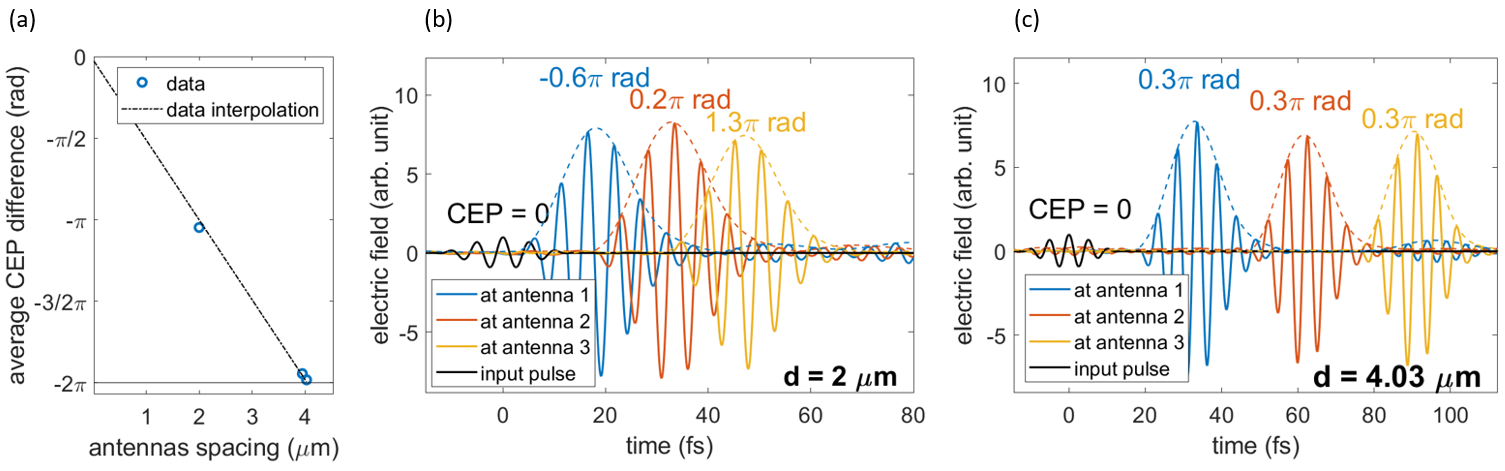}
\caption{(a) Calculated average difference in CEP between bow-tie antennas as a function of the spacing $d$. This calculation was used to optimize the spacing $d$, such that CEP possessed the same value at each antenna. With $d = 0\,\mu$m, no CEP shift occurs. (b) Field at antennas tips for spacing $d = 2\,\mu$m, which shows different CEP values at each antenna. (c) Field at antennas tips for spacing $d = 4.03\,\mu$m, which ensures a constant CEP at each antenna.}
\label{fig_CEP}
\end{figure*}

In order to find the optimal spacing between antennas along the waveguide, we simulated the device response using an ideal $\cos^2$-shaped pulse with an average pulse duration of 10~fs FWHM at a central wavelength of 1.55 $\mu$m before moving on to more realistic supercontinuum pulses that one would likely use for CEP detection in an integrated photonic system.  We evaluated the CEP value of the $\cos^2$-shaped pulse at each antenna along the waveguide, and figure \ref{fig_CEP}(a) shows the calculated average CEP difference from one bow-tie nanoantenna to the next as a function of the spacing $d$. In the limit of $d \approx 0\,\mu$m, no CEP shift occurs. We find that the optimal spacing to have the same CEP value at each nanoantenna for the assumed waveguide and device geometry was $d=4.03~\mu$m.

To better demonstrate the effect of antenna spacing, we show in figures \ref{fig_CEP}(b) and (c), the time-domain waveform at the tips of each device when using an incorrect and correct spacing.  The simulated time-domain response of the plasmonic bow-tie nanoantennas was obtained by an inverse Fourier transform of the frequency-domain response\cite{QNN:yujia, QNN:vanishing, QNN:putnam}.  Figure \ref{fig_CEP}(b) shows the enhanced pulses at the tips of three bow-tie antennas with spacing $d = 2\,\mu$m, given the chosen injected pulse with initial CEP $= 0$ rad. For this spacing value, the waveform CEP shifts by $\approx \pi$ rad. On the other hand, with spacing $d = 4.03\,\mu$m (Figure \ref{fig_CEP}(c)), it is clear that the CEP value is the same at each antenna resulting in identical enhanced field waveforms at the tip surface.
We note that there is a $0.3\pi$ CEP offset that is the same for each tip as a result of the resonant plasmonic coupling between each antenna and the waveguide mode. Since this shift is maintained throughout all of the antennas, it is possible to extrapolate the injected pulses CEP value.
We note that the time-domain field enhancement of $\approx$ 7-8$\times$ was also observed.

To this point we have ignored nonlinear effects within the waveguide.  Given the large peak intensities potentially involved, it is important to consider the contribution from third-order nonlinear phase accumulation in the waveguide, as this would also impact the CEP and potentially the waveform shape. To gauge the importance of such third-order nonlinearity, we calculated it by the B-integral. For a Si\textsubscript{3}N\textsubscript{4} waveguide having a length of $\approx 40\,\mu$m (long enough to contain nine bow-tie nanoantennas at the correct spacing), an accumulated shift of just $\frac{\pi}{5}$ was estimated at the end of the waveguide. This small phase shift indicates that over this length, self phase modulation is not significant. Furthermore, we point out that this shift is distributed along the waveguide length and could be taken into account in the device design by slightly tuning the spacing between the antennas.  Furthermore, possible defects and uncertainties in the device materials and dimensions may also affect the CEP slippage in real devices, and will need to be considered during fabrication.

\section{Results \& Discussion}
\label{results}

In this section, we present the device response to few-cycle pulses that could be realistically excited via on-chip supercontinuum generation.
In sub-section \ref{cep_sens_signal_subsection} we analyze the CEP-sensitive photocurrent after interaction with up to nine devices in series. In sub-section \ref{SNR_subsection} we estimate the SNR to shot noise current and then in sub-section \ref{losses_subsection} we investigate the device loss contributions.

\subsection{CEP-Sensitive Signal}
\label{cep_sens_signal_subsection}

\begin{figure*}[!t]
  \centering
  \includegraphics[width=1\linewidth]{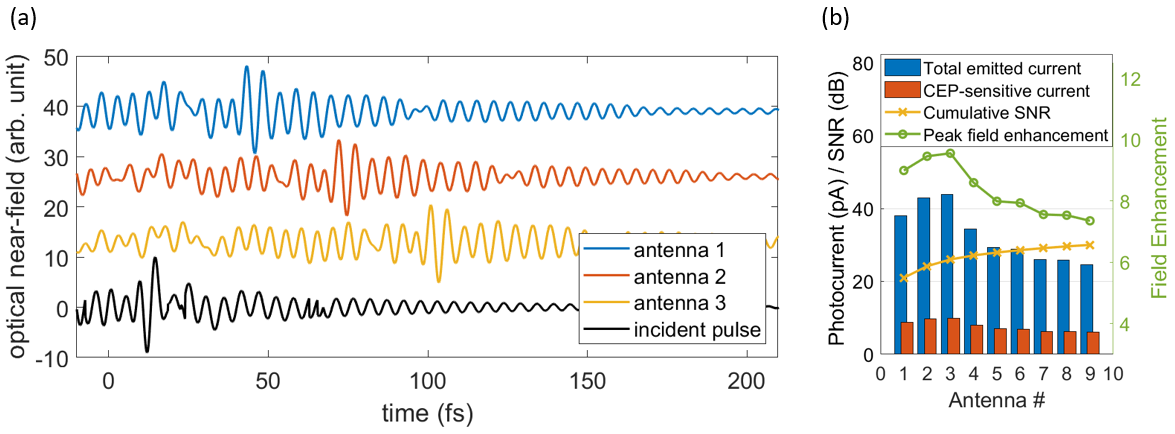}
\caption{(a) Time-domain response of the device to a supercontinuum pulse train. The input pulse and pulses enhanced at the tips of the three bow-tie nanoantennas, according to the field enhancement in Figure \ref{fig_FE}, are displayed. The represented input waveform is not plotted on the same scale as the enhanced waveforms (the input waveform is enhanced $\approx6\times$ in amplitude). Note that the enhanced waveforms possess the same CEP value at each antenna tip as the antennas are placed at the correct spacing $d$. (b) Bar plot of the current contribution of each antenna, peak field enhancement at the antennas tips and cumulative SNR at RBW = 50 kHz, for a total of nine antennas.
}
\label{fig_real_pulse}
\end{figure*}

The CEP-sensitivity of the nanoantennas increases dramatically with reduced duration of the enhanced driving pulse~\cite{QNN:yujia}.  Few-cycle pulses can be generated directly within integrated photonics platforms using supercontinuum generation \cite{paper:singh_pulse, paper:pulses_dudley}.  In order to estimate a realistic few-cycle pulse for interaction with the devices, we simulated on-chip supercontinuum generation within a SiN waveguide that could be directly coupled to the waveguide containing the device structures.  For the calculations, we assumed that an input pulse having a duration of 100 fs and energy of 100 pJ was injected into a SiN waveguide having a cross section of 1100 nm width and 800 nm height. Due to strong self phase modulation and the anomalous dispersion of the waveguide, the output pulse duration is significantly compressed relative to the input. We assume that this fully compressed pulse, which occurs right before the soliton fission point and is shown as the black trace in Fig.~\ref{fig_real_pulse}(a), is then directly injected into the waveguide containing the nanoantenna devices.

Figure \ref{fig_real_pulse}(a) shows the resulting input pulse and enhanced pulses at the tips of the first three bow-tie nanoantennas. Note that the input waveform is not plotted on the same scale as the enhanced waveforms for visualization purposes (the input waveform is amplified $\approx6\times$ in amplitude). Note that here, as before, with the correctly designed spacing $d = 4.03$~$\mu$m, the enhanced waveforms in Figure \ref{fig_real_pulse}(a) possess approximately the same CEP value at each antenna tip (\textit{i.e.} the enhanced waveforms are almost identical).

The injected and enhanced pulses present high pulse compression, although with a strong pedestal, and are centered at a wavelength of $1.55\,\mu$m (see Appendix \ref{methods} for pulse train spectrum). Assuming a maximum pulse energy of 100~pJ, this results in a peak input field at the center of the waveguide of $\approx$ 2.6 GV/m, and peak fields at the surface of the antennas of $\approx$ 15 GV/m. Under these conditions, we calculate a Keldysh parameter $\gamma \approx 0.6$, which means that the emitted photocurrent can be approximated using a quasi-static Fowler-Nordheim tunneling rate~\cite{paper:keldysh, paper:yalunin, QNN:putnam}. We then used the enhanced waveforms at the nanoantennas tips to calculate the emitted photocurrent using the simplified quasi-static Fowler-Nordheim tunnelling model \cite{paper:FN_simplified}.

The potential barrier chosen for the gold-air interfaces was 5.1 eV, which has shown to provide a good description of tunneling rates in past experimental work~\cite{QNN:putnam, QNN:vanishing} (see Appendix \ref{methods} for further information).
The effective emission surface area at the antennas tips, used for photoemission estimation, was calculated by comparing the experimental CEP-sensitive photocurrent given by the free-space bow-tie nanoantennas array in \cite{QNN:yujia} to the numerical CEP-sensitive current density that was obtained by simulating the electromagnetic and time-domain response of the same array (see Appendix \ref{methods} for further information).  Using this approach, we estimated the emission surface area to be $\approx 36\,\text{nm}^2$.


In order to obtain a stronger signal and SNR, we considered up to nine antennas connected in series. Figure \ref{fig_real_pulse}(b) shows the peak FE for nine antennas placed along the waveguide.  The FE at the antennas tips generally decreases moving from tip to tip along the waveguide, mainly due to power scattering and absorption at the metal structures. It shows a slight increase at the tips of antennas 2 and 3, that can be ascribed to the excitation of cavity resonances in the regions between consecutive antennas.
Figure \ref{fig_real_pulse}(b) also shows the current contribution of each antenna assuming a repetition rate of 100 MHz. 

\begin{figure}[!h]
  \centering
  \includegraphics[width=0.9\linewidth]{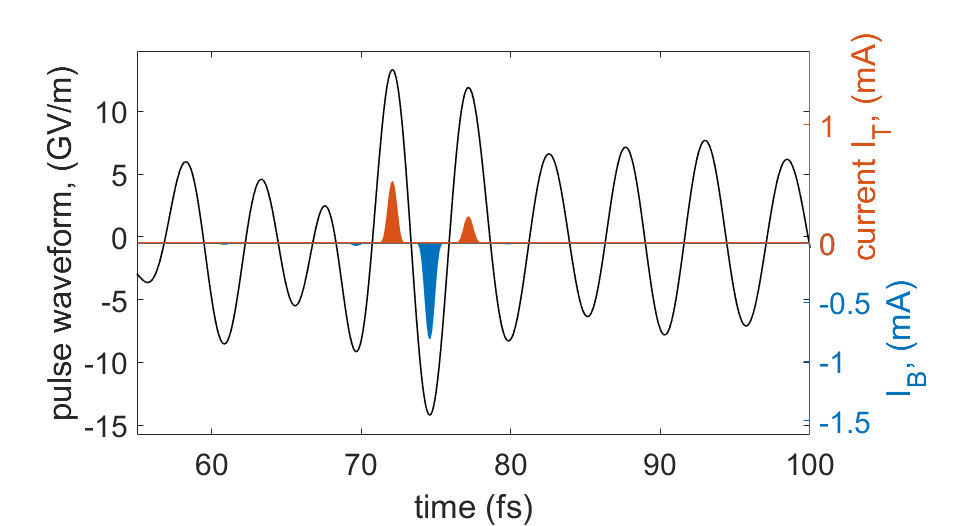}
\caption{Enhanced waveform at the tip of antenna 1 and photocurrent emitted from the top and bottom nanotriangle tip.  Current is delivered in sub-cycle bursts due to the high nonlinearity and field-dependence of the quasi-static tunneling process.}
\label{fig_real_pulse_2}
\end{figure}

The nine bow-tie antennas generate a net CEP-sensitive current of $I_{\text{cep}}\approx 70$~pA (corresponding to $I_{\text{cep}}\approx $ 8 pA per bow-tie antenna) and a total emitted current (\textit{i.e.} the sum of the current from every nanotriangle including current that cannot be detected due to the cancellation of photocurrents of opposite directions)  of $I_{\text{emit}}\approx 300$~pA. The CEP-sensitivity, defined as $I_{\text{cep}} / I_{\text{emit}}$, was found to be $\approx 0.2$ for each nanoantenna. 
This value is much higher than the CEP-sensitivities obtained in past works \cite{QNN:yujia, QNN:vanishing, QNN:putnam}, which were measured to be between $10^{-5}$ and $10^{-3}$ for 10 fs incident pulses at a central wavelength of 1177 nm. This increase is primarily due to the reduced number of cycles per pulse of the enhanced tip fields, given by an extremely short time duration and a larger central wavelength of the injected pulses. For instance, when examining the current density emitted from the tip of antenna 1 in Figure \ref{fig_real_pulse_2}, we see that the emission is limited to just the three central half-cycles, explaining the increased CEP-sensitivity.  As explained in prior work, CEP decreases significantly as the number of contributing half-cycles increases~\cite{QNN:drew, QNN:vanishing} (also see further analysis provided in Appendix \ref{appendix_2}).  

We note that due to the supercontinuum generation process, the calculated pulse has a rather long pedestal.  For high enough pulse energies, these pedestal fields can start to generate nontrivial amounts of photocurrent.  
When the total charge emission from the pedestal becomes significant relative to that from the peak fields, the CEP-sensitivity starts to reduce, decreasing the device performance. For the conditions and waveforms considered here, we estimate that this reduction in CEP-sensitivity due to pedestal emission occurs for peak fields $\gtrapprox$ 5 GV/m.  

\subsection{SNR Estimation}
\label{SNR_subsection}

Experiments in \cite{QNN:yujia} determined that the dominant source of noise for such nanoantenna CEP detectors is shot noise resulting from the total emitted photocurrent $I_{\text{emit}}$. 
The shot noise current scales as $\sqrt{2q\,\Delta f\, I_{\text{emit}}}$, with $q$ the electron charge and $\Delta f$ the resolution bandwidth. The resulting SNR is proportional to the square of the CEP-sensitivity and to the total emitted photocurrent $I_{\text{emit}}$. We calculate that each antenna contributes to the total SNR and that an SNR of 30 dB at RBW = 50 kHz should be obtainable from a linear array of just nine bow-tie nanoantennas, as shown in Figure \ref{fig_real_pulse}(b). 

This SNR is mainly due to the high CEP-sensitivity, allowed by extremely compressed pulses. We find that the SNR calculated here is comparable to the one from state-of-the-art self-referencing techniques ($\approx 30$ dB at RBW of tens to hundreds of kHz \cite{paper:okawachi_gaeta_LN_integr_combs, ceo_stab_Klenner:16, ceo_stab_Okawachi:18}) and suitable for CEP measurement and control. Hence this device could provide an alternative and compact route to conventional $f$-$2f$ interferometers for optical-frequency-comb stabilization.

\subsection{Analysis of Optical Losses}
\label{losses_subsection}

\begin{figure}[!h]
  \centering
  \includegraphics[width=0.9\linewidth]{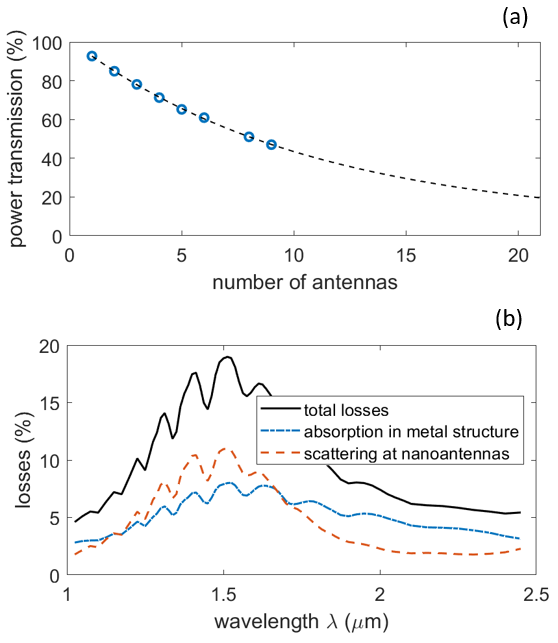}
\caption{(a) Loss of power transmitted through the waveguide at $\lambda = 1.55\,\mu$m for devices with an increasing number of connected antennas. Dashed lines indicate the prediction of the data trend, given by an exponential interpolation of the data. (b) Power losses as a function of injected wavelength. Losses are primarily attributed to power absorption at the metal structure and scattering by the antennas.}
\label{fig_losses}
\end{figure}

Figure \ref{fig_losses}(a) shows the loss of power transmitted through the waveguide at $\lambda = 1.55\,\mu$m for devices with an increasing number of connected antennas.
The device presents a loss for each added antenna of about $5.7\%$ of the total input power.

This graph shows that power is theoretically drained out of the device almost completely after 20-30 bow-tie nanoantennas. Furthermore, the steep decrease in transmitted power suggests that only the first antennas of a device would generate a strong-enough CEP-sensitive signal. Nevertheless, even though antenna 8 receives only $\approx 40-50\,\%$ of the total input power, the enhanced field still maintains $\approx 80\%$ of the value of FE at the initial antenna, as seen in Figure \ref{fig_real_pulse}(b), and is able to generate a significant amount of photocurrent. 
We note that power losses limit the number of devices that can be operated in series for an individual waveguide and hence the total generated SNR per waveguide. Nonetheless, if the power of the input pulse train is strong enough, multiple waveguides with bow-tie devices could be excited in parallel in order to achieve larger values of SNR.

Figure \ref{fig_losses}(b) shows the losses as a function of wavelength for a device with a total of three connected antennas.
The device simulation shows power losses mainly due to absorption at the metal structure (i.e., nanoantennas and connecting wires) and scattering by the antennas. Each of these two contributions provide a maximum loss value of 5-10$\%$ of the input power at $\lambda \approx 1.55\,\mu$m, to a total maximum loss of around 15-20$\%$. Losses are mainly caused by the field interacting with metal structures and hence follow the trend of field enhancement at the antenna tips with a peak at around $1.55\,\mu$m in wavelength and weak modulation caused by the excitation of cavity resonances. 

\section{Conclusion}

In this work, we have designed electrically-connected nanoantenna arrays on a Si$_3$N$_4$ waveguide structure for CEP detection within integrated photonic platforms.  In the design process, we matched the device plasmonic resonance to the central wavelength of the injected pulses by tuning the nanoantenna geometry enabling time-domain field enhancements of the injected ultrashort pulses on the order of 7-8$\times$ at the tips of the antennas. Moreover, we showed how proper antenna spacing can accommodate for CEP slippage along the waveguide ensuring that the CEP-sensitive signal from each antenna adds in phase, which is critical for achieving adequate CEP-sensitive signal and SNR.  

Our work showed that such an approach is attractive as it offers: (1) direct time-domain CEP detection within photonic platforms, (2) a reduction to three orders of magnitude in foot-print from $\approx$ mm to $\approx \mu$m \cite{paper:okawachi_gaeta_LN_integr_combs, ceo_stab_Klenner:16}, and (3) no need of external detectors for conversion to electronic signal. With this simple structure we presented an alternative to integrated $f$-$2f$ interferometers for CEP measurement and control. The calculated SNR of 30 dB at RBW = 50 kHz is comparable to those obtained from state-of-the-art waveguide-integrated self-referencing techniques ($\approx 30$ dB at RBW of tens to hundreds of kHz \cite{paper:okawachi_gaeta_LN_integr_combs, ceo_stab_Klenner:16, ceo_stab_Okawachi:18}). Furthermore, the designs we present here could be extended to waveguide-integrated tools for petahertz-scale optical field sampling \cite{paper:mina}.  As such, this work represents the first steps towards fully-integrated light-wave-based PHz electronics \cite{paper:perspect_petahz_electr, paper:krausz_stockman_atto_metr, paper:goulielmakis, paper:lee_petahertz_1, paper:lee_petahertz_2}.


\section*{Acknowledgment}

This material is based upon work supported by the Air Force Office of Scientific Research under award numbers FA9550-19-1-0065 and FA9550-18-1-0436.  We thank Karl K. Berggren and Marco Colangelo for their scientific discussion and edits to the manuscript.

\bibliographystyle{IEEEtran}
\bibliography{Bibliography.bib}

\appendices

\section{Methods / Model Setup}
\label{methods}
The device electromagnetic response was simulated with a finite element method electromagnetic solver (\textit{COMSOL Multiphysics}). The modeled device consists of a total of three gold bow-tie nanoantennas placed on top of a Si$_3$N$_4$-core waveguide. Each nanoantenna consists of a pair of nanotriangles placed in a bow-tie-like disposition. The waveguide has a Si$_3$N$_4$ core with rectangular cross-section of $1\,\mu \text{m} \times 0.8\,\mu \text{m}$ and a SiO$_2$ bottom cladding up to the top surface of the core. We studied the antennas exposed to air environment.
The nanoantennas are interconnected by 40-nm-wide gold nanowires and the whole gold structure (connecting wires and nanotriangles) is 20 nm thick. In the design, a 2-nm-thick chromium adhesion layer was assumed between the waveguide core and the gold nanostructures.

The modeled nanoantennas and connecting wire geometries were initially taken from the SEM images of a fabricated nanoantennas array of a previous work of the group \cite{QNN:yujia}. They were successively adjusted to match simulation necessities, i.e. tuning the antennas plasmonic resonance to the input pulses central wavelength. In particular, the final design value for the nanotriangles altitude is $\approx$ 250~nm, the base is $\approx$ 190~nm, their vertices are rounded with a radius of curvature of 5 nm and the gap between the nanotriangles in each bow-tie antenna is $\approx$ 50 nm.
The spacing between bow-tie antennas $d$ along the waveguide was set to $d=4.03\,\mu$m.

Frequency-domain simulations were performed over a broad wavelength range [1.025$\,\mu$m $-$ 2.5$\,\mu$m].
Optical properties of gold and chromium were taken from \cite{paper:johnson}, properties for Si$_3$N$_4$ from \cite{paper:luke}
and for SiO$_2$ from \cite{paper:sio2_1, paper:sio2_2}. The input-wave port was defined as the front cross-section of the whole waveguide (cladding and core) and of the connecting wires. Perfectly matched layers were added all around  the simulation domain to absorb outgoing electromagnetic waves and model semi-infinite cladding.
The plane-wave light was polarized as a transverse magnetic (TM) mode, with $(E_x,E_y,H_z)$ non-zero components, according to the axis convention shown in Figure \ref{fig_device}(a). $E_y$, which is polarized along the bow-tie long-axis to excite the plasmonic mode, resulted the main electric field component.

The field enhancement is defined as the optical field at the nanotriangle tip near the nano-gap, averaged over the curved surface described by the tip radius of curvature and the gold thickness, divided by a field reference value $E_0$. The reference value is calculated as the maximum optical field value in the center of the waveguide core at the entrance of the waveguide, calculated for a simple waveguide (with no metal structure) that is the same dimension as the one considered in this work. This reference value varies as a function of wavelength.
Power losses were found to be dominated by two main contributions: power absorbed at the metal structure (both Au and Cr) and power scattered by the nanoantennas. The latter is calculated by detecting the power that crosses the external boundaries of the waveguide cladding.

\begin{figure}[!h]
  \centering
  \includegraphics[width=0.8\linewidth]{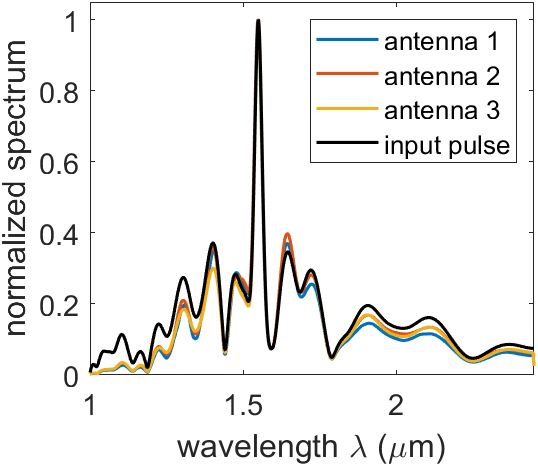}
\caption{Normalized spectrum of the considered supercontinuum pulses, showing the central wavelength at $\lambda = 1.55\,\mu$m.}
\label{fig_SC_spectrum}
\end{figure}

For the time-domain response and the photocurrent estimation, two different input pulses were considered: (1) an ideal test cos$^2$-shaped pulse train with an average pulse duration of 10~fs FWHM at a central wavelength of $1.55\,\mu$m, a maximum pulse energy of 100~pJ and a repetition rate of 100~MHz, (2) an ultrafast supercontinuum pulse train having a central wavelength of $1.55\,\mu$m, a maximum pulse energy of 100~pJ and a repetition rate of 100~MHz (see the description in the manuscript body for details regarding the simulation of the supercontinuum pulses). Figure \ref{fig_SC_spectrum} shows the injected and enhanced supercontinuum spectra at the first three antenna tips.
The spectrum of the pulses was obtained by a Fourier transform. These spectra were then filtered by the frequency response of the device at each of the antennas tips and the resulting time-domain response was obtained by an inverse Fourier transform. 

The photocurrent emission was estimated by the simplified quasi-static Fowler-Nordheim tunnelling model.
The considered emission surface at the antennas tips was estimated to be $\approx 36\,\text{nm}^2$ and was calculated by comparing the experimental CEP-sensitive photocurrent given by the free-space bow-tie nanoantennas array in \cite{QNN:yujia} to the numerical CEP-sensitive current density that was calculated by simulating the electromagnetic and time-domain response of the same array. \cite{QNN:yujia} reported an average CEP-sensitive photocurrent of 1.3 pA per bow-tie antenna and the simulations resulted in a CEP-sensitive current density of $0.036$ pA/nm\textsuperscript{2}.

In order to predict the photocurrent emission of up to nine antennas on a single waveguide, the antenna spacing was reduced to 1 $\mu$m to reduce computational overhead. This was justified as we found that the spacing parameter does not greatly affect the power dissipation along the waveguide as most power is lost through scattering and absorption from each antenna.  We further confirmed this by manually altering the spacing and observing little change in loss.  The CEP at each antenna was then manually adjusted to be the same.


\section{Influence of pulse peak field and pulse compression}
\label{appendix_2}

In this appendix we study the influence of the input pulse features on device performance.

First, we calculated the CEP-sensitive current and SNR for different injected peak field strengths, in the range $[0.5\,\,\text{GV/m}- 5\,\,\text{GV/m}]$, which corresponds to around $[10\,\text{pJ}-260\,\text{pJ}]$ in energy, for supercontinuum pulses as the ones employed in previous sections. This calculation is important in order to evaluate possible emission from the field pedestal of these pulses and to understand its influence on the device sensitivity.

\begin{figure}[!h]
  \centering
  \includegraphics[width=0.9\linewidth]{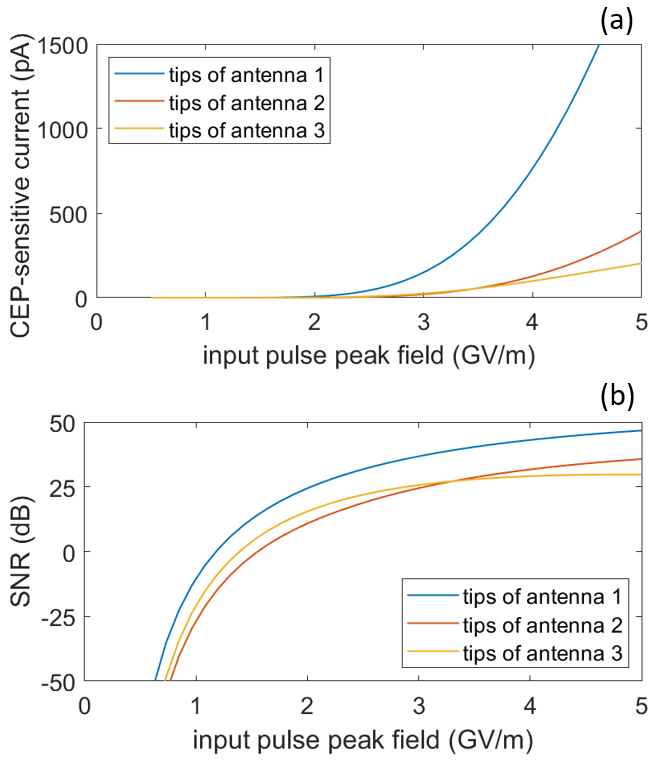}
\caption{(a) CEP-sensitive current as a function of the input pulse peak field, for each of the three considered antennas. (b) CEP-sensitive signal to shot noise ratio SNR at RBW = 50 kHz as a function of the input pulse peak field, for each of the considered antennas.}

\label{fig_pulse_variation_field}
\end{figure}

Figures \ref{fig_pulse_variation_field}(a) and \ref{fig_pulse_variation_field}(b) show the CEP-sensitive current and the relative shot noise SNR at RBW = 50 kHz respectively, as a function of the injected pulse peak field before enhancement, for each of the three considered antennas. 
Figure \ref{fig_pulse_variation_field}(a) shows an exponential increase in CEP-sensitive current for stronger and stronger peak fields. This is an expected behaviour, since stronger fields result in larger photoemission and the largest contribution to it is given by the field at the pulse peak, due to its non-linear characteristics.
On the other hand, for strong fields, a higher tunneling probability arises, causing possible emission from the pulse pedestal.
This undesired emission leads to a lower contribution of CEP-sensitive current with respect to total emitted current, so a drop in device CEP-sensitivity.
This drop can be seen in the resulting SNR, in Figure \ref{fig_pulse_variation_field}(b). SNR is proportional to the square of the CEP-sensitivity and to the total emitted photocurrent $I_{\text{emit}}$.
Whereas the total emitted current increases exponentially, CEP sensitivity drops, and the SNR saturates to 30-45 dB.

In both figures, a clear difference is observed in the response from each antenna. This difference is a result of small differences in the FE at the antennas along the waveguide, which lead to different enhanced fields and hence different emission values. 

For CEP detection, two factors need to be taken into account: (1) a large enough CEP-sensitive photocurrent is needed to be able to detect variation of CEP, and (2) a large enough SNR is required to overcome noise.
We calculate that with a linear array of 10-15 bow-tie nanoantennas, both large enough CEP-sensitive current and SNR of 30 dB at RBW = 50 kHz can be obtained for pulses with peak field larger than 2.5 GV/m.

\begin{figure}[!h]
  \centering
  \includegraphics[width=0.9\linewidth]{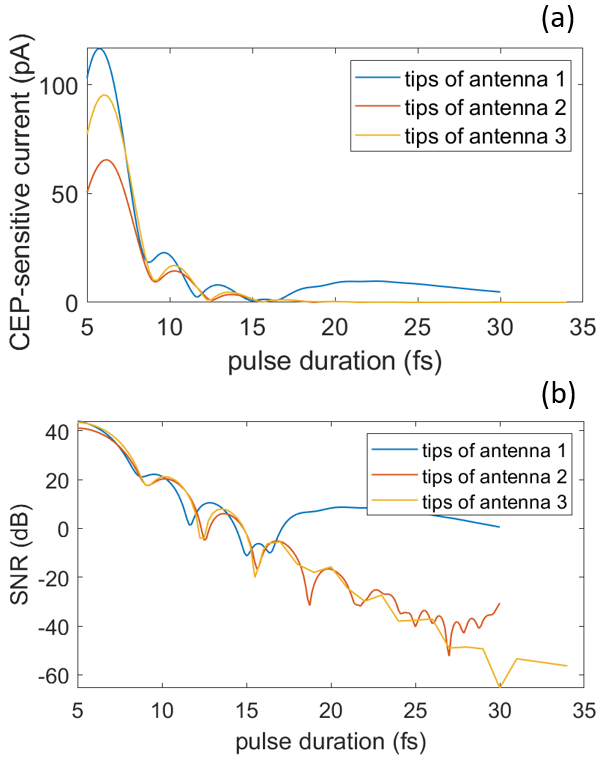}
\caption{(a) CEP-sensitive photocurrent emitted by each of the antennas tips as a function of the pulse duration in fs, for the optimal spacing value $d$. (b) CEP-sensitive signal to shot noise ratio SNR at RBW = 3 kHz as a function of pulse duration, for each of the considered antennas.}
\label{fig_pulse_variation_time}
\end{figure}

Finally, we study the impact of the time duration of the incident pulses on the device performance, which demonstrates the crucial role of highly compressed input pulses. For this study, we simplified the input pulses to ideal transform-limited pulses similar to those used in section \ref{results} for the integration design testing. We calculate the CEP-sensitive current and SNR for different pulse time duration, in the range $[5\,\text{fs}-30\,\text{fs}]$.

Figure \ref{fig_pulse_variation_time}(a) shows the CEP-sensitive photocurrent emitted by each of the antennas tips as a function of the pulse duration in fs. As seen in the Supplementary Information of \cite{QNN:yujia}, a short pulse duration, which corresponds to an extremely low number of optical-field cycles in the waveform, leads to strong CEP-sensitivity and can generate a large quantity of CEP-sensitive photocurrent from each antenna. This dependence is due to the fact that fewer cycles in a pulse lead to the presence of fewer side field lobes contributing to the cancellation of the dependence on CEP, which is given by the central field peaks of the pulse. In contrast, as pulse duration gets higher and higher, the side field lobes increase in number and in their contribution to the total emitted current and hence the CEP-sensitive current decreases. The total emitted photocurrent increases for higher pulse duration, because more sub-cycles are strong enough to drive photocurrent emission.

Figure \ref{fig_pulse_variation_time}(b) displays the CEP-sensitive signal to shot noise ratio SNR at a resolution bandwidth of 3 kHz, as a function of pulse duration. For very short pulse duration, an extremely large SNR of $\approx 40$ dB can be obtained from each bow-tie antenna. For larger pulse duration SNR decreases due to lower CEP-sensitivity, then a larger number of antennas are required to obtain the SNR needed for CEP detection. The oscillating behaviour of CEP-sensitive photocurrent and SNR can be ascribed to the fact that as the pulse duration increases, more and more pulse lobes are considered and as each of them become increasingly relevant, a dip in CEP-sensitive current is observable.
From these results, we calculate that with a linear array of 10-15 bow-tie nanoantennas an SNR of 30 dB at a resolution bandwidth of 3 kHz can be obtained only for pulses with duration of 10 fs or smaller.

\end{document}